\def\half{\frac{1}{2}}
\documentclass{iopart}
\usepackage{graphicx}
\begin{document}
\hspace*{2.5 in}CUQM-113, HEPHY-PUB 813/05\\

\hspace*{2.5 in}math-ph/0602059\\

\vspace*{0.6 in}
\title{Gravitating semirelativistic $N$-boson systems}
\author{Richard L. Hall}
\address{Department of Mathematics and Statistics, Concordia University,
1455 de Maisonneuve Boulevard West, Montr\'eal,
Qu\'ebec, Canada H3G 1M8}
\author{Wolfgang Lucha}
\address{Institute for High Energy Physics, Austrian Academy of Sciences,
Nikolsdorfergasse 18, A-1050 Vienna, Austria}
\eads{\mailto {rhall@mathstat.concordia.ca}, \mailto {wolfgang.lucha@oeaw.ac.at}}
\begin{abstract}Analytic energy bounds for $N$-boson systems governed by
semirelativistic Hamiltonians of the~form $H=\sum_{i=1}^N({\bf
p}_i^2+m^2)^{1/2}-\sum_{1=i<j}^Nv/r_{ij},$ with $v>0,$ are derived by use of
Jacobi relative coordinates. For gravity $v=c/N,$ these bounds are
substantially tighter than earlier bounds and they are shown to coincide with
known results in the nonrelativistic limit.
\end{abstract}

\noindent\pacs{03.65.Ge, 03.65.Pm}
\vskip0.2in
\maketitle
\section{Introduction: the $N$-body problem}
One-body Hamiltonians $H$
composed of the relativistic expression $\sqrt{{\bf p}^2+m^2}$ for the
kinetic energy of particles of mass $m$ and momentum ${\bf p}$ and of a
coordinate-dependent static interaction potential $V({\bf r}),$ defined as
operator sum$$H=\sqrt{{\bf p}^2+m^2}+V({\bf r}),$$provide a simple but very
efficient tool for the description of relativistically
moving particles~\cite{BSE,SE,Lieb96,Lucha:O,Lucha:D,Lucha04:TWR}.
 They have been used, for
instance, for the description of hadrons as bound states of quarks
\cite{Lucha91,Lucha92}. One of the advantages of this kind of
semirelativistic treatment is that its generalization to the many-body
problem is straightforward to formulate. A semirelativistic Hamiltonian for a
system of $N$ identical particles interacting by pair potentials $V(r_{ij})$
is given by
\begin{equation}H=\sum_{i=1}^N\sqrt{{\bf p}_i^2+m^2}+\sum_{1=i<j}^NV(r_{ij}).
\label{Eq:H}\end{equation}We shall use the notational simplification
$p\equiv|{\bf p}|,$ $r\equiv|{\bf r}|$, or $r_{ij}\equiv|{\bf r}_i-{\bf
r}_j|,$ whenever no ambiguity is introduced by doing so. Many approaches to
such many-body problems for identical particles employ the very powerful
constraint of permutation symmetry to generate their reduction to a two-body
problem with a Hamiltonian ${\mathcal H}$ whose spectrum is used to
approximate the many-body energy eigenvalues or to generate a lower energy
bound. This reduction may be effected in various ways, which leads to the
problem of finding the most effective reduced problem, the one which would
provide the {\it highest\/} lower bound. In this paper, we use Jacobi
relative coordinates which introduce technical difficulties but yield a lower
bound that is much improved over previous results. We have already achieved a
similar improvement for the harmonic-oscillator potential \cite{Hall03} and
for convex transformations of it \cite{Hall04}. Here, we show that one can
derive tight $N$-body energy bounds for the (more physically important) case
of the gravitational potential: $V(r)=-v/r,$ $v>0.$ There is a long history
of attention to the corresponding nonrelativistic problem, dating back at
least to 1967 \cite{Hall67,Hall92}. Also the semirelativistic case has been
discussed before, notably by Lieb and Thirring \cite{Lieb84}, by Lieb and Yau
\cite{Lieb87}, as well as by Martin and Roy \cite{Martin89}.

In their rigorous investigation \cite{Lieb87} of Chandrasekhar's theory of
the stellar collapse and its formal connection to quantum physics, Lieb and
Yau were able to demonstrate exactly that in the simultaneous limit
$N\to\infty$ and $v\to0$ of particle number $N$ and gravitational coupling
constant $v$ such that their product $Nv$ is kept fixed (at an arbitrary
value below some critical value) the Schr\"odinger equation reduces to the
``semiclassical'' approximation represented by a Hartree-type equation for
the density of the bosons. In particular, they succeeded in proving the
convergence of the lowest quantum energy, $E,$ defined as the infimum of the
spectrum of the Hamiltonian $H$ given by Eq.~(\ref{Eq:H}), to the
corresponding semiclassical minimum Hartree energy. Interestingly, precisely
such kind of behavior in the limit $N\to\infty$ has been conjectured already
before in Ref.~\cite{Lieb84}.

The study in Ref.~\cite{Martin89} shows that for the one-body case with a
gravitational potential a lower bound to the energy spectrum may be expressed
in terms of the lowest energy of a Schr\"odinger operator with a Kratzer
potential, which is a Coulomb potential ``spiked'' with an additional term of
the form $A/r^2.$ In spite of the complications of relative coordinates but
with boson permutation symmetry in the individual-particle coordinates, we
may take advantage of this result to construct improved lower energy bounds
for systems comprised of identical particles with integer~spin.

The plan of this paper is as follows. After recalling, in Sec.~\ref{Sec:CLB},
existing Coulomb one-body lower energy bounds, and analyzing, in
Sec.~\ref{Sec:SLB}, the simple ($N/2$) $N$-body~lower energy bound, we
reformulate, in Sec.~\ref{Sec:JRC}, the $N$-particle problem in relative
coordinates. This allows us to derive, in Sec.~\ref{Sec:ILB}, an improved
lower energy bound. To this~result we adjoin, in Sec.~\ref{Sec:GUB}, a
variational (Rayleigh--Ritz) upper energy bound obtained with the help of a
scale-optimized Gaussian trial wave function. We summarize our results in
Sec.~\ref{Sec:RAC}, and inspect, in particular, their large-$N$ limit.

\section{One-body lower energy bounds for the relativistic Coulomb problem}
\label{Sec:CLB}
The case of a one-particle Hamiltonian with a Coulomb
potential $V(r)=-v/r$ has been extensively investigated. We know from the
pioneering work of Herbst \cite{Herbst} that the Hamiltonian is self-adjoint
for $v\le\frac{1}{2}$ and has a Friedrichs extension up to the critical value
$v=2/\pi.$ Herbst obtains the following lower bound to the ground-state
energy $E$:$$E\ge m\sqrt{1-\left(\frac{\pi v}{2}\right)^2},\quad 0\le
v<\frac{2}{\pi}.$$For the smaller range $0\le v<\frac{1}{2}$ of the Coulomb
coupling $v,$ this result was strengthened by Martin and Roy
\cite{Martin89}~to\begin{equation}E\ge
m\sqrt{\frac{1+\displaystyle\sqrt{1-(2v)^2}}{2}},\quad 0\le v<\frac{1}{2}.
\label{Eq:MR}\end{equation}This energy bound was found by considering the
squares of the kinetic-energy and the potential-energy terms and by relating
certain expectations to the previously studied exact solution of the
Klein--Gordon Schr\"odinger equation. Let us now briefly re-derive this same
result, emphasizing that this lower bound is, in fact, generated by the
lowest energy of a Schr\"odinger operator \cite{Schroed}; this latter problem
had earlier been analyzed by Kratzer and others~\cite{Kratzer,Landau,Fluegge,Morales,Hall98,Znojil,Hooydonk}.

 We suppose that
the exact normalized ground state of $H$ is $\psi$ so that
$(p^2+m^2)^{1/2}\psi=(E+v/r)\psi.$ By considering the equality between the
squares of these equal vectors we obtain, upon introducing a quantity $\ell$
by $\ell(\ell+1)=-v^2,$\begin{eqnarray}
E^2-m^2&=&\left(\psi,\left(p^2-\frac{2Ev}{r}-\frac{v^2}{r^2}\right)\psi\right)
\nonumber\\&\ge&-\frac{E^2v^2}{(\ell+1)^2};\label{Eq:E2m2LB}\end{eqnarray}the
inequality on the right-hand side of (\ref{Eq:E2m2LB}) arises from the
variational principle applied to the Kratzer Hamiltonian:$$H_K=
p^2-\frac{2Ev}{r}+\frac{\ell(\ell+1)}{r^2}=p^2-\frac{2Ev}{r}-\frac{v^2}{r^2}.$$
The well-known expression for the bottom, for given $\ell,$ of the hydrogen
energy spectrum remains valid for negative non-integer $\ell$ provided
$|\ell|<\frac{1}{2},$ which is equivalent to the constraint $v<\frac{1}{2}.$
By solving for $\ell$ in terms of the coupling parameter $v,$ we recover the
Martin--Roy lower~bound~(\ref{Eq:MR}).

\section{``Simple'' (or $N/2$) lower energy bound for general $N$-body
problems}\label{Sec:SLB}
This rather simple lower energy bound is not limited
to the gravitational pair potential, and it allows us to prove that a given
$N$-body Hamiltonian is bounded from below. Applying the same reasoning to
the (soluble) Schr\"odinger harmonic-oscillator problem, defined by the
Hamiltonian $$H=\sum_{i=1}^N{\bf p}_i^2+\sum_{1=i<j}^Nr_{ij}^2,$$with the
exact ground-state energy $E=3(N-1)\sqrt{N}$, one gets only $E/\sqrt{2}$
whereas a general lower bound based on Jacobi relative coordinates yields
indeed the exact energy for this potential \cite{Hall95}. A lower bound to
the ground-state energy of the Hamiltonian (\ref{Eq:H}) is provided by the
bottom ${\mathcal E}_{N/2}$ of the spectrum of the one-body Hamiltonian operator
\begin{equation}{\mathcal H}_{N/2}=N\left[\sqrt{p^2+m^2}+\frac{N-1}{2}V(r)\right],
\label{Eq:calH2}\end{equation}since boson permutation symmetry of the exact
$N$-body normalized wave function $\Psi$ implies $(\Psi,H\Psi)=(\Psi,h\Psi),$
where the operator $h$ is a two-body Hamiltonian given~by$$h=\frac{N}{2}
\left[\sqrt{{\bf p}_1^2+m^2}+\sqrt{{\bf p}_2^2+m^2}+(N-1)V(r_{12})\right].$$
Changing the coordinates of this two-particle problem to ${\bf r}={\bf
r}_1-{\bf r}_2$ and ${\bf R}={\bf r}_1+{\bf r}_2,$ the
individual momenta are given, in terms of the corresponding total and
relative momenta variables, by ${\bf p}_{1,2}={\bf P}\pm{\bf p}.$ Using the
lemma of Ref.~\cite{Hall03} to remove the center-of-mass momentum term ${\bf P},$
the reduced two-body Hamiltonian $h$ becomes ${\mathcal H}_{N/2}.$ If we now apply
this simple bound to the gravitational problem, $V(r) = -v/r,$ then we find,
from (\ref{Eq:calH2}) and the one-body lower bound (\ref{Eq:MR}),
\begin{equation}E\ge Nm\sqrt{\frac{1+\displaystyle\sqrt{1-(N-1)^2v^2}}{2}}
,\quad(N-1)v<1.\label{Eq:SLB}\end{equation}

\section{$N$-particle problems in terms of Jacobi relative coordinates}
\label{Sec:JRC}
 Our Hamiltonian $H$ does not have the kinetic energy of the
$N$-particle system's center-of-mass removed.~Thus,~its eigenvectors are
subject to two fundamental symmetries: translation invariance, and boson
permutation symmetry (in the individual-particle coordinates $\{{\bf
r}_1,{\bf r}_2,\dots,{\bf r}_N\}.$ 
Jacobi relative coordinates may be defined with the aid of an orthogonal
matrix $B$ transforming old ($\{{\bf r}_i\}$) to new
($\{\mbox{\boldmath{$\rho$}}_i\}$) coordinates by
$[\mbox{\boldmath{$\rho$}}]=B[{\bf r}].$ The first row of $B,$ with~all entries
$B_{1i}=1/\sqrt{N},$ defines a center-of-mass variable
$\mbox{\boldmath{$\rho$}}_1,$ its second a pair distance
$\mbox{\boldmath{$\rho$}}_2=({\bf r}_1-{\bf r}_2)/\sqrt{2},$ and the $k$th
row ($k\ge2$) first has $k-1$ entries $B_{ki}=1/\sqrt{k(k-1)},$ the $k$th
entry $B_{kk}=-\sqrt{(k-1)/k},$ and zero for all remaining entries. The
momenta $\{\mbox{\boldmath{$\pi$}}_i\}$ conjugate to the
$\{\mbox{\boldmath{$\rho$}}_i\}$ read $[\mbox{\boldmath{$\pi$}}]=(B^{-1})^{\rm
T}[{\bf p}]=B[{\bf p}].$ Now, for an $N$-boson problem with an attractive
potential $V(r),$ let $\Psi(\mbox{\boldmath{$\rho$}}_2,
\mbox{\boldmath{$\rho$}}_3,\dots,\mbox{\boldmath{$\rho$}}_N)$ be the (still
to be found) normalized ground-state eigenfunction corresponding to the
lowest energy $E.$ Boson symmetry is a powerful constraint that greatly
reduces the complexity of this problem. Although a non-Gaussian wave function
is not necessarily symmetric in the Jacobi coordinates, we do have the
remarkable $N$-representability expressions \cite{Hall03}
(Appendix A here) for $i,j > 1$
\begin{eqnarray}
\left(\Psi,(\mbox{\boldmath{$\rho$}}_i\cdot\mbox{\boldmath{$\rho$}}_j)\Psi\right)&=\delta_{ij}\left(\Psi,\mbox{\boldmath{$\rho$}}_2^2\Psi\right)\\    \left(\Psi,(\mbox{\boldmath{$\pi$}}_i\cdot \mbox{\boldmath{$\pi$}}_j)\Psi\right)&=\delta_{ij}\left(\Psi,\mbox{\boldmath{$\pi$}}_2^2\Psi\right).
\label{Eq:Nrep}
\end{eqnarray}
We now choose to use ${\bf p}_{N}$ and ${\bf p}_{N-1}$ and we obtain the reduction
$$E=(\Psi,H\Psi)=\left(\Psi,\left[\frac{N}{2}\sqrt{{\bf p}_N^2+m^2}+\frac{N}{2}\sqrt{{\bf p}_{N-1}^2+m^2}+\gamma
V(|{\bf r}_{N-1}-{\bf r}_N|)\right]\Psi\right),$$
where $\gamma=\frac{1}{2}N(N-1).$
We first note that the last row of $B$ (which defines ${\bf p}_N$) is given by 
$$[a, a, a, \dots, a, -(N-1)a],\quad {\rm where}\quad a = 1/\sqrt{N(N-1)}.$$
Now we express the equation for $E$ in terms of new coordinates defined by the following relations
$${\bf r} = {\bf r}_{N-1}-{\bf r}_N = \alpha \mbox{\boldmath {$\rho$}}_{N}-\beta\mbox{\boldmath {$\rho$}}_{N-1}, \quad r = |{\bf r}|,$$
$$\left[ \begin{array}{c}
{\bf R}\\
{\bf r}\end{array} \right] = 
\left[\begin{array}{cc}
\beta & \alpha\\
\alpha & -\beta \end{array}\right]
\left[ \begin{array}{c}
\mbox{\boldmath {$\rho$}}_N\\
\mbox{\boldmath {$\rho$}}_{N-1}\end{array} \right],
\quad 
\left[ \begin{array}{c}
{\bf P}\\
{\bf p}\end{array} \right] = 
\half\left[\begin{array}{cc}
\beta & \alpha\\
\alpha & -\beta \end{array}\right]
\left[ \begin{array}{c}
\mbox{\boldmath {$\pi$}}_N\\
\mbox{\boldmath {$\pi$}}_{N-1}\end{array} \right],$$
where 
$$\alpha = \sqrt{\frac{N}{N-1}}> 1,\quad\beta = \sqrt{\frac{N-2}{N-1}}< 1,\quad \delta = \sqrt{\frac{N-2}{N}} < 1.$$
Consequently
$$\alpha^2 + \beta^2 = 2,\quad a^2 + \beta^2 = \frac{1}{\alpha^2} = \lambda = \frac{N-1}{N},$$
and
$$(N-1)a = \alpha^{-1},\quad Na = \alpha,\quad \delta = \frac{\beta}{\alpha},
\quad 1+\delta^2 = 2\lambda.$$
We note that because of the boson symmetry, and after removal of $\pi_1,$ we have generally that
$$\langle f(|{\bf p}_{N-1}|)\rangle = \langle f(|{\bf p}-\delta {\bf P}|)\rangle = \langle f(|{\bf p}+\delta {\bf P}|)\rangle = \langle f(|{\bf p}_{N}|)\rangle,$$
where $f(p)$ is any appropriate kinetic-energy function.
The expression for the energy now has the form
$$E = \left\langle\frac{N}{2}\sqrt{\left({\bf p}+\delta {\bf P}\right)^2 + m^2}+ \frac{N}{2}\sqrt{\left({\bf p}-\delta {\bf P}\right)^2 + m^2}+\gamma V(r)\right\rangle,$$
or, equivalently,
\begin{equation}
E = \langle {\mathcal H}\rangle,\quad {\rm where}\quad {\mathcal H} = N\sqrt{\left({\bf p}+\delta {\bf P}\right)^2 + m^2}+\gamma V(r).
\label{Eq:Hm}
\end{equation}
The Hamiltonian ${\mathcal H}$ is bounded below by the simple bound ${\mathcal H}_{N/2}$ of Eq.~(\ref{Eq:calH2}). This result is proved in Appendix B. We now consider the eigenequation ${\mathcal H}\psi = {\mathcal E}\psi,$ where $\psi({\bf r},{\bf R})$ retains some of the boson symmetry implications of the full wave function.  What we need are immediate consequences of Eq.~(\ref{Eq:Nrep}), namely $\langle {\bf p}^2\rangle = \langle {\bf P}^2\rangle,$
 $\langle {\bf p}\cdot{\bf P}\rangle = 0,$ and hence
\begin{equation}
\langle ({\bf p}+\delta {\bf P})^2\rangle = \langle p^2 + \delta^2 P^2\rangle = (1+\delta^2)\langle p^2\rangle = 2\lambda \langle p^2\rangle.
\label{Eq:bsym}
\end{equation}

\section{Improved lower energy bound}\label{Sec:ILB}
\begin{figure}[htbp]\centering\includegraphics[width=12cm]{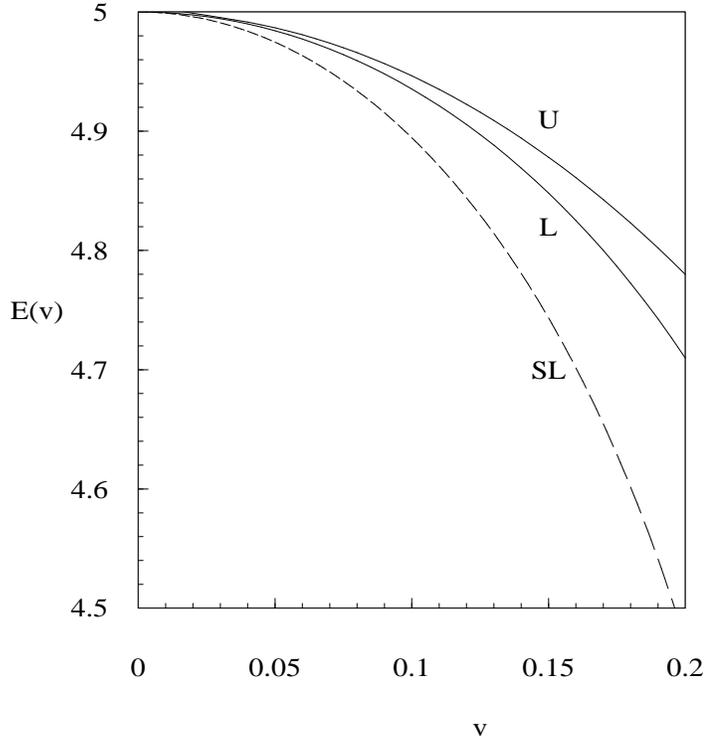}
\caption{Energy bounds $E(v)$ (in dimensionless units) for the
semirelativistic $N$-body problem with $N=5$ and mass $m=1$, as functions of
the dimensionless coupling parameter $v.$ The figure shows the simple lower
bound (SL), the improved lower bound (L), and a scale-optimized Gaussian
upper bound (U).}\label{Fig:E(v)}\end{figure}

The lower bound that
improves the $N/2$ bound (\ref{Eq:SLB}) and forms our main result is
specifically for the attractive pair potential $V(r)=-v/r,$ $v>0$. Suppose
that $\psi({\bf r}, {\bf R})$ is the
exact lowest eigenfunction of the Hamiltonian ${\mathcal H}$ in
(\ref{Eq:Hm}). We look in a restricted domain ${\mathcal D}$ in the
Hilbert space $L^2(R^6),$ one that keeps some of the original boson
symmetry, namely, we assume Eq.~(\ref{Eq:bsym}).
This allows us to apply the same reasoning as we did for the
one-body problem because the kinetic term is squared and thus the final form
of the expectation values involves only the conjugate variables $\{{\bf r}, {\bf p}\}.$
 Let ${\mathcal E}$ be the minimum of $(\psi,{\mathcal H}\psi)$ corresponding to normalized
$\psi$ satisfying (\ref{Eq:bsym}). ${\mathcal E}$ is a lower bound to $E,$
${\mathcal E}\le E,$ because, if $\Psi$ is the exact normalized $N$-body wave
function, then
$$E=(\Psi,H\Psi)=(\Psi,{\mathcal H}\Psi)\ge(\psi,{\mathcal H}\psi)={\mathcal E}.$$
The eigenvalue equation of ${\mathcal H}$, ${\mathcal H}\psi={\mathcal E}\psi,$ explicitly
reads
$$N\sqrt{({\bf p}+\delta {\bf P})^2 +m^2}~\psi=\left({\mathcal E}
+\frac{\gamma v}{r}\right)\psi.$$
Squaring these equal vectors and use of (\ref{Eq:bsym}), and also the scaling change $\{r, p\}\rightarrow \{2r, p/2\}$,  implies
$${\mathcal E}^2-N^2m^2\ge\gamma\inf_{\psi\in{\mathcal
D}}\left(\psi,\left( p^2-\frac{v{\mathcal E}}{r}-\frac{\gamma
v^2}{4r^2}\right)\psi\right).$$By comparing this with the one-body case
(\ref{Eq:MR}), we obtain\begin{equation}E\ge{\mathcal E}\ge
Nm\sqrt{\frac{1+\displaystyle\sqrt{1-\gamma v^2}}{2}},\quad\gamma
v^2<1.\label{Eq:ILB}\end{equation}Figure~\ref{Fig:E(v)} illustrates the
improvement by the lower bound (\ref{Eq:ILB}) over the previously available
simple lower bound (\ref{Eq:SLB}).

\section{Gaussian upper energy bound}\label{Sec:GUB}
In order to find an upper
bound, we follow Ref.~\cite{Hall04}~and use a Gaussian wave function, which
we write in the
form$$\Phi(\mbox{\boldmath{$\rho$}}_2,\mbox{\boldmath{$\rho$}}_3,\dots,
\mbox{\boldmath{$\rho$}}_N)
=C\exp\left(-\frac{\alpha}{2}\sum_{i=2}^N\mbox{\boldmath{$\rho$}}_i^2\right),$$
where $\alpha>0,$ while $C$ guarantees the normalization of $\Phi.$
The boson symmetry of the trial function allows us to write $E \leq E_G = \left(\Phi,H\Phi\right),$ where we have
 
\begin{eqnarray}
E_G &=  \left(\Phi,\left[N\sqrt{{\bf p}_N^2 + m^2} + \gamma V(|{\bf r}_1-{\bf r}_2|)\right]\Phi\right)\\
&= \left(\Phi,\left[N\sqrt{\lambda \mbox{\boldmath {$\pi$}}_N^2 + m^2} +\gamma V(|\sqrt{2}\mbox{\boldmath{$\rho$}}_2|)\right]\Phi\right).
\end{eqnarray}

The symmetry of the Gaussian function in the relative coordinates and the factoring property allow us to replace 
$\mbox{\boldmath{$\pi$}}_N$ by $\mbox{\boldmath{$\pi$}}_2 \equiv {\bf p},$ and, setting ${\bf r}\equiv\mbox{\boldmath{$\rho$}}_2$, we find explicitly
$$E\le E_G = N\left(\phi,\sqrt{\lambda
p^2+m^2}\phi\right)+\gamma\left(\phi,V(\sqrt{2}r)\phi\right),$$
where $\phi(r)=(\alpha/\pi)^{3/4}\exp(-\frac{1}{2}\alpha r^2).$ The kinetic-energy
expectation value may be expressed in terms of the modified Bessel function of
the second kind $K_{1}(x)$ \cite{Abramowitz}. Evaluating the integral, for
convenience, in momentum space, yields$$\left(\phi,\sqrt{\lambda
p^2+m^2}\phi\right)=\frac{2m}{\mu}\sqrt{\frac{2}{\pi}}g\left(\frac{\mu^2}{4}
\right),$$where $\mu=m\{2N/[(N-1)\alpha]\}^{1/2},$ while $g(x)$ is defined~by
$$g(x)=x\exp(x)K_1(x)=\int_{-\infty}^{+\infty}dt\,t^2\sqrt{2x+t^2}\exp(-t^2).$$
The potential-energy expectation value reads for the case of a gravitational
pair-interaction potential $V(r)=-v/r$:$$\left(\phi,V(\sqrt{2}r)\phi\right)
=-\frac{Nmv}{\mu}\sqrt{\frac{2}{\pi\gamma}}.$$With the expectation values in
this form, we may regard the parameter $\mu$ as a variational parameter. We
arrive~at\begin{equation}E\le Nm\sqrt{\frac{2}{\pi}}\min_{\mu>0}
\left[\frac{2g(\mu^2/4)-\sqrt{\gamma}v}{\mu}\right],\quad N\ge2.
\label{Eq:GUB}\end{equation}The analytical fact that the function
$[2g(\mu^2/4)-a]/\mu$ has a minimum only if $a<2$ entails the constraint
$v<2/\sqrt{\gamma}.$ Figure~\ref{Fig:E(v)} confronts the two lower bounds
(\ref{Eq:SLB}) and (\ref{Eq:ILB}) with this ``scale-optimized Gaussian''
variational upper~bound.

\begin{figure}[htbp]\centering\includegraphics[width=12cm]{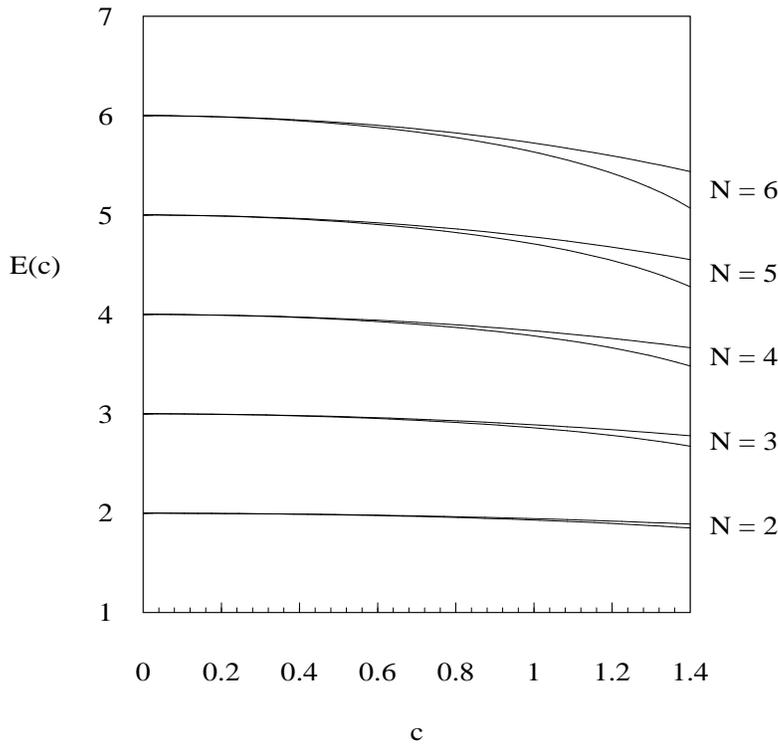}
\caption{Upper and lower bounds to the ground-state energy $E(c)$ (in
dimensionless units) of the gravitational $N$-particle problem,
$N=2,3,\dots,6,$ with mass $m=1$, as functions of~the dimensionless coupling
parameter $c.$}\label{Fig:E(c)}\end{figure}

\section{Results and conclusion}\label{Sec:RAC}
Any consideration of an
arbitrarily large number, $N,$ of self-gravitating bosons, that is, the
inspection of the limit $N\to\infty,$ obliges us to force the gravitational
coupling $v$ to decrease for increasing $N$ in some appropriate manner, in
order to avoid the (``relativistic'') collapse $E\to-\infty$ for
$N\to\infty.$ Keeping $v$ fixed, inevitably implies collapse \cite{Martin88}.
Consequently, following Lieb and Yau \cite{Lieb87}, let $v$ diminish, with
increasing $N,$ according to $v=c/N,$ where $c$ is some {\it constant\/}
gravitational interaction parameter. In this case the term $(N-1)^2v^2$ in
the simple lower bound (\ref{Eq:SLB}) will be replaced by $\lambda^2c^2$
(with $c<1$), the improved lower bound (\ref{Eq:ILB}) is modified by
replacing $\gamma v^2$ by $\frac{1}{2}\lambda c^2$ (with $c<\sqrt{2}$), and
in the Gaussian upper bound (\ref{Eq:GUB}) we must substitute
$\sqrt{\lambda/2}c$ (with $c<2\sqrt{2}$) for $\sqrt{\gamma}v.$
Figure~\ref{Fig:E(c)} shows resulting energy curves $E(c)$ for
$N=2,3,\dots,6$ and $m=1.$ For the lower bounds we require for the
dimensionless parameter $c<\sqrt{2};$ the upper bounds are valid, for all
$N$, if $c<2\sqrt{2}.$

Some special cases are of interest. For small coupling~$c,$ both bounds are
parabolic in shape. Explicitly, they read

\begin{equation}
Nm\left(1-\frac{\lambda
c^2}{16}\right)\le E(c)\le Nm\left(1-\frac{\lambda c^2}{6\pi}\right).
\label{Eq:Bounds}
\end{equation}
These energy bounds coincide with previous findings for the corresponding
nonrelativistic problem; cf.\ Eq.~(4.2) of Ref.~\cite{Hall92}. Meanwhile, for
all couplings $c<\sqrt{2}$ (such that both constraints on $c$ are satisfied),
in the limit $N\uparrow\infty$~we have $\lambda\rightarrow1,$ and the general
results (\ref{Eq:ILB}) and (\ref{Eq:GUB}) provide bounds immediately to
$E(c)/Nm$ for arbitrarily large $N.$

The energy spectrum of a system of nonrelativistic identical
bosons experiencing any attractive pair interaction is bounded from below by
the lowest energy of a specially scaled one-body problem \cite{Hall67}. For the
harmonic oscillator, this bound yields the exact energy; such results are
facilitated by use of (orthogonal) relative coordinates. Tight energy bounds
have recently been obtained \cite{Hall04} for semirelativistic problems with
oscillator pair interactions. In this analysis we have constructed both lower
and upper bounds to the lowest energy of a semirelativistic system of
identical gravitating bosons. Our bounds reduce to their nonrelativistic
counterparts in the limit of weak coupling.
 \section*{Acknowledgement}
One of us (RLH) gratefully acknowledges both partial
financial support of his research under Grant No.~GP3438 from~the Natural
Sciences and Engineering Research Council of Canada and hospitality of the
Institute for High Energy Physics of the Austrian Academy of Sciences in
Vienna.
 \section*{Appendix A}
 For definiteness, we consider the relative momenta $\mbox{{\boldmath $\pi$}}_i$ and $\mbox{{\boldmath $\pi$}}_j$ with $i, j >1.$ Since $[\mbox{{\boldmath $\pi$}}] = B[\mbox{{\boldmath $p$}}],$ we have
$$\mbox{{\boldmath $\pi$}}_i\cdot\mbox{{\boldmath $\pi$}}_j = \left(\sum_k B_{ik}{\bf p}_k\right)\cdot\left(\sum_k B_{jk}{\bf p}_k\right) = \sum_{k}B_{ik}B_{jk}\left({\bf p}_k^2\right) + \sum_{k\ne l}B_{ik}B_{jl}\left({\bf p}_k\cdot {\bf p}_l\right).\eqno{\rm (A1)}$$
By using the boson symmetry of the wave function we have
$$\langle \mbox{{\boldmath $\pi$}}_i\cdot\mbox{{\boldmath $\pi$}}_j\rangle =\left(\sum_{k}B_{ik}B_{jk}\right)\langle {\bf p}_1^2\rangle + \left(\sum_{k\ne l}B_{ik}B_{jl}\right)\langle {\bf p}_1\cdot{\bf p}_2\rangle.\eqno{\rm (A2)}$$
But orthogonality of the rows of $B$ to the first row tells us
$$0 = \sum_{k}B_{ik} = \left(\sum_{k}B_{ik}\right)\left(\sum_{l}B_{jl}\right) = \left(\sum_{k}B_{ik}B_{jk}\right) + \left(\sum_{k\ne l}B_{ik}B_{jl}\right).\eqno{\rm (A3)}$$
If $ i = j$, then the orthogonality of the matrix $B$ tells us that
$$\sum_{k}B_{ik}^2 = 1= -\sum_{k\ne l}B_{ik}B_{il}.$$
Meanwhile, if $i \ne j,$ we know that the corresponding rows of $B$ are orthogonal; hence, in this case, $\sum_{k}B_{ik}B_{jk} = 0;$ thus, both of the coefficients in (A2) are zero.  Hence we conclude for all $i,j > 1$
$$\langle \mbox{{\boldmath $\pi$}}_i\cdot\mbox{{\boldmath $\pi$}}_j\rangle = \delta_{ij}\langle \mbox{{\boldmath $\pi$}}_2^2\rangle = \delta_{ij}\left(\langle {\bf p}_1^2\rangle - \langle {\bf p}_1\cdot {\bf p}_2\rangle\right).\eqno{\rm (A4)}$$
The proof for the relative coordinates $\{\mbox{{\boldmath $\rho$}}_i\}$ is identical.
 \section*{Appendix B}
The vectors ${\bf P}$ and ${\bf p}$ (with $\mbox{{\boldmath $\pi$}}_1$ removed) define a plane; we let ${\bf k}$ be a unit vector perpendicular to this plane.  Then, for example, we have
$$\left(m^2 + ({\bf p}+\delta {\bf P})^2\right)^{\half} = \left|m{\bf k} + {\bf p} + \delta {\bf P}\right|\eqno{\rm (B1)}$$
In this notation the expectation of the kinetic-energy term in ${\mathcal H}$, defined in Eq.~(\ref{Eq:Hm}), reads
$$\left\langle\left|m{\bf k} + {\bf p} + \delta {\bf P}\right|\right\rangle = 
\left\langle\left|m{\bf k} + {\bf p} - \delta {\bf P}\right|\right\rangle.\eqno{\rm (B2)}$$
Now, we may write 
$$m{\bf k} + {\bf p} = \half[(m{\bf k} + {\bf p} + \delta {\bf P})+(m{\bf k} + {\bf p} - \delta {\bf P})].\eqno{\rm (B3)}$$
The triangle inequality then tells us
$$\left|m{\bf k}+ {\bf p}\right| \le \half\left|m{\bf k} + {\bf p} + \delta {\bf P}\right| + \half\left|m{\bf k} + {\bf p} - \delta {\bf P}\right|.\eqno{\rm (B4)}$$
If we now look at mean values, we see from (B2) that
$$\left\langle\left|m{\bf k} + {\bf p}\right|\right\rangle \le \left\langle\left|m{\bf k}+ {\bf p}+\delta {\bf P}\right|\right\rangle,\eqno{\rm (B5)}$$
or, equivalently
$$\left\langle\left(m^2 + p^2\right)^{\half}\right\rangle \le \left\langle\left(m^2 + ({\bf p}+\delta {\bf P})^{2}\right)^{\half}\right\rangle.\eqno{\rm (B6)}$$
Thus we conclude that $\langle H \rangle = \langle {\mathcal H} \rangle\ge \langle {\mathcal H}_{N/2} \rangle,$ where the reduced one-body Hamiltonian ${\mathcal H}_{N/2}$ is given by
$${\mathcal H}_{N/2} =N\left(m^2 + p^2\right)^{\half} + \gamma V(r).\eqno{\rm (B7)}$$
We note the spectral inequality  ${\mathcal H} \ge {\mathcal H}_{N/2}$ with equality only when $N = 2.$  Thus $ E = \langle H\rangle = \langle{\mathcal H}\rangle \geq \langle{\mathcal H}_{N/2}\rangle.$
\vspace{1.5cm}
\end{document}